\documentclass[a4paper,12pt,3p]{elsarticle}

\newcommand{\rmd}{\mathrm{d}}

\begin{document}

\title{Agent based reasoning for the non-linear stochastic models of long-range memory}

\author{A. Kononovicius}
\ead{aleksejus.kononovicius@gmail.com}

\author{V. Gontis\corref{cor}}
\ead{vygintas@gontis.eu}

\cortext[cor]{Corresponding author}

\address{Vilnius University, Institute of Theoretical Physics and Astronomy, A. Go\v{s}tauto 12, 01108 Vilnius, Lithuania}

\begin{abstract}
We extend Kirman's model by introducing variable event time scale. The proposed flexible time scale is equivalent to the variable trading activity observed in financial markets. Stochastic version of the extended Kirman's agent based model is compared to the non-linear stochastic models of long-range memory in financial markets. Agent based model providing matching macroscopic description serves as a microscopic reasoning of the earlier proposed stochastic model exhibiting power law statistics.
\end{abstract}


\maketitle

\section{Introduction}

Computational modeling of complex systems \cite{waldrop1992} has
become a rapidly developing method of doing science
\cite{axelrod1998}. This approach is indispensable when
microscopic nature of interactions is unambiguously defined.
Nevertheless in the prevailing majority of social systems
microscopic nature of interactions has to be deduced from
the macroscopic behavior of the whole system.

There are two major types of financial market models -
microscopic, most usually agent based, and macroscopic, usually
based on stochastic calculus as most common choice in
phenomenological description of complex systems. Agent based
models (ex. \cite{cristelli2010, lux1999, kirman1993}) come from
the basic ideas about behavior inside the system. Resulting models
can be rather simple \cite{kirman1993} or not so \cite{lux1999}.
No matter the complexity of description there are still no agent
based models, which would be able to sufficiently mimic empirical
data \cite{cristelli2010}. In contrast, stochastic models (ex.
\cite{gontis2010Scyio,  reimann2011}) are being derived directly
from the empirical data, thus reproducing observed statistical
features, so-called stylized facts \cite{cont2001, bouchaud2004}.
The drawback of stochastic models is the lack of direct insights
into the microscopic nature of replicated dynamics. Bridging
between these two very different approaches one could
propose model successful in both - theory and practice.

Top-down approach, starting from stochastic and moving towards agent based models, seems to be a very formidable task, as macro-behavior of complex system can not be understood as simple superposition of varying micro-behaviors. In case of sophisticated agent based models \cite{cristelli2010} bottom-up approach provides too many opportunities. But there is a wide selection of rather simple agent based models (ex. \cite{kirman1993}), whose stochastic treatment can be directly obtained from the microscopic description \cite{alfarano2005}.

In this contribution we consider the opportunity to generalize
Kirman's ant colony model \cite{kirman1993} with the intention to
modify microscopic approach to the financial market \cite{alfarano2005} reproducing
main stylized facts of this complex system. In Section
\ref{sec:kirman} we will briefly introduce Kirman's ant colony
model, propose its generalization and derive stochastic model for
population dynamics. Next, in Section \ref{sec:stochasticRet}, we
will introduce return, modulating return and propose single
stochastic model for the modulating return. In Section
\ref{sec:comparison} we compare generalized Kirman's model with
stochastic description of financial markets analyzed in
\cite{reimann2011, ruseckas2010}. Conclusion and discussion of
future prospects will be given in Section \ref{sec:conclusion}.

\section{Kirman's model and its stochastic treatment}
\label{sec:kirman}

In his seminal paper, \cite{kirman1993}, Kirman noticed that entomologists and economists observe similar patterns in rather different systems.

Kirman credits entomologists Deneubourg and Pastels to be the first ones to observe ant colony with two identical food sources available (for the original papers see references in \cite{kirman1993}, while relevant research is also available in more recent paper \cite{detrain2006}). They observed that in such case majority of ants still tend to use only a single food source at any given time. Though the other food source is not completely neglected as switches to the previously overlooked food source were observed despite the fact that food sources remain identical.

Interestingly enough human crowd behavior tends to be quite similar, at least in statistical sense. There are observations that majority of people tend to choose more popular product, than less popular, despite both being of a similar quality. The article \cite{kirman1993} also cites numerous works, which speculate that herding behavior might be related to the endogenous fluctuations of asset price.

Taking discussed empirical observations into account Kirman proposed a Markovian chain with the following transition probabilities:
\begin{eqnarray}
p (X \rightarrow X+1) &=& (N-X) (\sigma_1 + h X) \Delta t ,\label{eq:pPlus}\\
p (X \rightarrow X-1) &=& X [\sigma_2 + h (N-X)] \Delta t ,\label{eq:pMinus}
\end{eqnarray}
where $N$ is a total number of agents (i.e. ants, traders or etc.) in the system (i.e. colony, market or etc.), $X$ is a number of agents choosing one option (i.e. food source, trading strategy), while $N-X$ use another option, $\Delta t$ is a very short time step, for which $p (X \rightarrow X+1) + p (X \rightarrow X-1) \leq 1$ holds. In the above transition probabilities $h$ terms describe herding behavior, as intrinsic property of agents themselves. While $\sigma_i$ terms describe individual transitions, which are assumed to be motivated by the attractiveness of the other option.

Kirman also gathered empirical evidence that the aforementioned
behavioral patterns are observed in ant colonies with very
different populations - from tens to millions of ants. This is
most probably true in socio-economic scenarios. Thus it would be
convenient to obtain a model description in the
continuous limit, i.e. unrelated to $N$. One
can assume that $N$ approaches infinity securing continuity of $x
= \frac{X}{N}$, in such case transition rates are expressed as
follows:
\begin{eqnarray}
\pi^{+} (x) &=& (1-x) \left[ \frac{\sigma_1}{N} + h x \right] , \label{eq:conpPlus} \\
\pi^{-} (x) &=& x \left[ \frac{\sigma_2}{N} + h (1-x) \right] .\label{eq:conpMinus}
\end{eqnarray}
Evidently the above are related to the original transition probabilities, Eqs. (\ref{eq:pPlus}) and (\ref{eq:pMinus}), as $ p (X \rightarrow X \pm 1) = N^2 \pi^{\pm}(x) \Delta t $.

Using one-step process formalism \cite{vanKampen1992} one can
compactly express Master equation via one-step operators,
$\mathbf{E}$ (increment) and $\mathbf{E}^{-1}$ (decrement):
\begin{equation}
\partial_t \omega(x,t) = N^2 \left\{ (\mathbf{E} -1) [ \pi^{-}(x) \omega(x,t) ] + (\mathbf{E}^{-1} -1) [ \pi^{+}(x) \omega(x,t) ] \right\} , \label{eq:master}
\end{equation}
where $\omega(x,t)$ is time-dependent probability density function. As one-step operators act on continuous functions one can expand them using Taylor series up to the second order terms. After doing so one would obtain Fokker-Plank equation:
\begin{equation}
\partial_t \omega(x,t) = - \partial_x [A(x) \omega(x,t)] + \frac{1}{2} \partial^2_x [B(x) \omega(x,t)] , \label{eq:FPeq}
\end{equation}
where in the limit of large $N$,
\begin{eqnarray}
A(x) &=& N (\pi^{+}(x)-\pi^{-}(x)) = \sigma_1 (1-x) - \sigma_2 x ,\\
B(x) &=& \pi^{+}(x)+\pi^{-}(x) = 2 h x (1-x) + \frac{\sigma_1}{N} (1-x) + \frac{\sigma_2}{N} x \approx 2 h x (1-x) .
\end{eqnarray}
This Fokker-Planck equation corresponds to the stochastic differential equation \cite{gardiner1997}:
\begin{equation}
\rmd x = \left[ \varepsilon_1 (1-x) - \varepsilon_2 x \right] \rmd
t_s + \sqrt{2 x (1-x)} \rmd W_s ,\label{eq:xsdedimless}
\end{equation}
where we have introduced dimensionless time scale, $t_s = h t$, and accordingly scaled the original model's parameters, $\varepsilon_i = \frac{\sigma_i}{h}$. In the above $W_s$ is appropriately scaled Wiener process. Eq. (\ref{eq:xsdedimless}), in non-dimesionless form, was originally derived in slightly different manner by Alfarano et al. in \cite{alfarano2005}.

\section{Stochastic Kirman's model for return in the financial markets}
\label{sec:stochasticRet}

As Kirman's model describes two state dynamics one must define two
types of agents acting inside the market in order to relate
Kirman's model to financial markets. Currently, the most
common choice is assuming that agents can be either
fundamentalists or noise traders \cite{cristelli2010}.

Fundamentalists are assumed to be long term investors who have
fundamental knowledge, which is quantified as fundamental price,
$P_f(t)$, of the traded stock. Thus their excess demand,
$D_f(t)$, is shaped by their long term expectations
\cite{alfarano2005}:
\begin{equation}
D_f(t) = N_f(t) \ln \frac{P_f(t)}{P(t)} ,
\end{equation}
where $N_f(t)$ is a number of fundamentalists inside the market and
$P(t)$ is a current market price. Being long term
investors fundamentalists assume that $P(t)$ will converge towards
$P_f(t)$ at least in a long run. Therefore if $P_f(t) > P(t)$,
fundamentalists will expect that $P(t)$ will grow in future and
consequently they will buy the stock ($D_f(t)>0$). In the opposite
case, $P_f(t) < P(t)$, they will expect decrease of $P(t)$ and
for this reason they will sell the stock ($D_f(t)<0$).

Noise traders on the other hand are short term investors who
estimate future price based on its recent movements. As there is a
wide selection of technical trading strategies, one can simply
assume that average noise traders demand is based on the
mood, $\xi(t)$, \cite{alfarano2005}:
\begin{equation}
D_c(t) = - r_0 N_c(t) \xi(t) ,
\end{equation}
where $D_c(t)$ is a total excess demand of
noise trader group, $N_c(t)$ is a number of noise traders inside the
market and $r_0$ can be seen as a relative noise trader
impact factor.

Price and, later, return can be introduced into the model by applying Walrassian scenario. One can assume that trading in the market occurs trough the market maker, who sets a fair price. As fair price should stabilize market, sum of all groups' excess demands must be equal to zero:
\begin{eqnarray}
D_f(t) + D_c(t) &=&  N_f(t) \ln \frac{P_f(t)}{P(t)}  - r_0 N_c(t) \xi(t) = 0 ,\\
P(t) &=& P_f(t) \exp \left[ r_0 \frac{N_c(t)}{N_f(t)} \xi(t) \right] ,
\end{eqnarray}
where without loosing generality one can assume that fundamental price remains constant, $P_f(t) = P_f$. Therefore return in selected time window $T$ is given by:
\begin{equation}
r(t)=\ln \frac{P(t)}{P(t-T)} = r_0 \left[ \frac{x(t)}{1-x(t)} \xi(t) -  \frac{x(t-T)}{1-x(t-T)} \xi(t-T) \right] , \label{eq:defrnonadiab}
\end{equation}
where we have set that $\frac{N_c(t)}{N}=x$ and $\frac{N_f(t)}{N}=1-x$. Alfarano et al. simplified the above by assuming that $x(t)$ is slower process than $\xi(t)$ \cite{alfarano2005}, obtaining adiabatic approximation of return:
\begin{equation}
r(t) = r_0 \frac{x(t)}{1-x(t)} \zeta(t) ,
\end{equation}
where $\zeta(t) = \xi(t)-\xi(t-T)$. If $\zeta(t)$ is modeled using spin-noise model, as in \cite{alfarano2005}, then $\frac{x(t)}{1-x(t)}$ can be seen as an absolute return.

Besides directly linking Kirman dynamics to the financial market scenario, we see a possibility to extend Kirman's model by using financial markets as an inspiration. It is known that trading activity (i.e. event rate) in financial markets is non-constant, while the original Kirman's transition probabilities, Eqs. (\ref{eq:pPlus}) and (\ref{eq:pMinus}), assume that agents meet (i.e. events occur) at a constant rate. As derivation of stochastic differential equation does not depend on the explicit form of $\pi^\pm(x)$ (see previous section), we can easily introduce our extension directly to the Eq. (\ref{eq:xsdedimless}):
\begin{equation}
\rmd x = \left[ \varepsilon_1 (1-x) - \frac{\varepsilon_2 x}{\tau(x)} \right] \rmd t_s + \sqrt{\frac{2 x (1-x)}{\tau(x)}} \rmd W_s ,\label{eq:genxsdedimless}
\end{equation}
where $\tau(x)$ adjusts time scale, of microscopic events, according to the current macroscopic system state, $x$. Note that $\varepsilon_1$ is not divided by $\tau(x)$ - it is a consequence of assumed rationality of fundamentalists, namely their individual behavior should not be influenced by the current trades.

Using Ito formula for variable substitution, which can be found in \cite{gardiner1997}, one can obtain stochastic differential equation for absolute return, $y=\frac{x(t)}{1-x(t)}$:
\begin{equation}
\rmd y = \left[\varepsilon_1 + y \frac{2-\varepsilon_2}{\tau(y)} \right] (1 + y) \rmd t_s + \sqrt{\frac{2 y}{\tau(y)}} ( 1 + y ) \rmd W_s . \label{eq:sdey}
\end{equation}
Note that absolute return, $y$ in the above, serves as a measure of volatility in the financial markets. It is known that volatility has long-range memory and correlates with trading activity and has probability density function with power law tail \cite{cont2001}.

\section{Comparison with selected stochastic models}
\label{sec:comparison}

In this contribution we study the case of $\tau(y)=y^{-\alpha}$. This selection might be backed by the fact that trading activity has positive correlation with volatility. In such case obtained stochastic differential equation, Eq. (\ref{eq:sdey}), in the limit of $y \gg 1$ is very similar to the stochastic models discussed in \cite{reimann2011, ruseckas2010}.

In series of papers, see \cite{ruseckas2010} for the most recent discussion and relevant past references, stochastic differential equation,
\begin{equation}
\rmd y = \left( \eta  - \frac{\lambda}{2} \right) y^{2 \eta -1} \rmd t + y^\eta \rmd W , \label{eq:JuliusBendra}
\end{equation}
was introduced as a class of stochastic differential equations
providing solutions with power law statistics. It was shown
that spectral density of stochastic variable defined by Eq.
(\ref{eq:JuliusBendra}) is a power law:
\begin{equation}
S(f) \sim \frac{1}{f^\beta}, \quad \beta=1+\frac{\lambda-3}{2(\eta-1)} .
\end{equation}
Stationary probability density function $p(y)$ of the
aforementioned variable $y$ is a power law as well
\begin{equation}
p(y) \sim y^{-\lambda}.
\end{equation}
Note that one has to introduce diffusion restriction, at least
from the side of small values, of $y$ in Eq.
(\ref{eq:JuliusBendra}).

\begin{figure}
    \centering
    \includegraphics[width=0.4\textwidth]{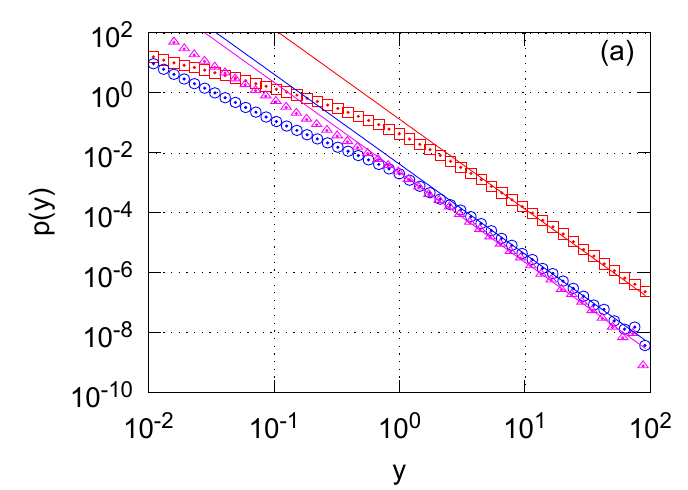}
    \hspace{0.1\textwidth}
    \includegraphics[width=0.4\textwidth]{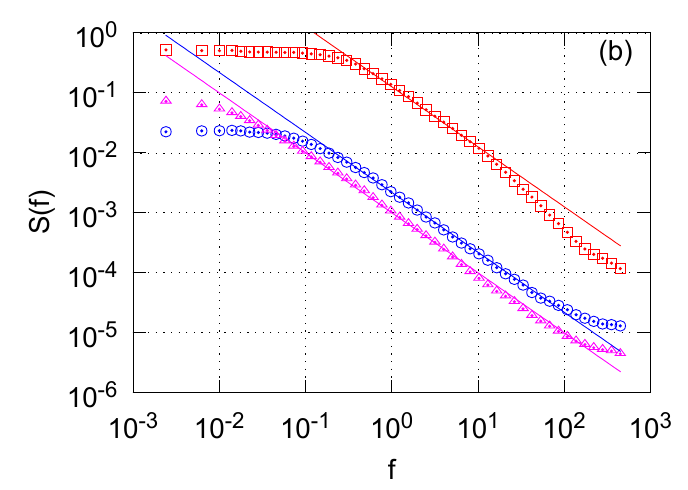}
    \caption{Numerically calculated PDF (a) and power spectral density (b) of $y$ defined by Eq. (\ref{eq:sdey}) in three distinct cases, $\alpha=0$ (red squares), $\alpha=1$ (blue circles) and $\alpha=2$ (magenta triangles). Other model parameters were set as follows: $\varepsilon_1=0$, $\varepsilon_2=2-\alpha$. Solid curves are analytical fits, Eqs. (\ref{eq:pdfJulius}) and (\ref{eq:specJulius}), for modelic results: (a) $\lambda=3$ (red squares, blue circles, magenta triangles), (b) $\beta=1$  (red squares, blue circles, magenta triangles).}
    \label{fig:compJulius}
\end{figure}

In the limit of large $y$, $y \gg 1$, we can consider only the highest powers in Eq. (\ref{eq:sdey}). In such case Eq. (\ref{eq:sdey}) becomes:
\begin{equation}
\rmd y = (2-\varepsilon_2) y^{2+\alpha} \rmd t_s + \sqrt{2 y^{3+\alpha}} \rmd W_s . \label{eq:sdeylarge}
\end{equation}
Direct comparison of Eqs. (\ref{eq:JuliusBendra}) and
(\ref{eq:sdeylarge}) gives:
\begin{eqnarray}
& \eta = \frac{3+\alpha}{2}, \label{eq:etadef}\\
& \lambda = \varepsilon_2 + \alpha +1 .
\end{eqnarray}
Consequently we expect stochastic variable $y$ defined by Eq. (\ref{eq:sdey}) to have power law stationary probability density function,
\begin{equation}
p(y) \sim y^{-\varepsilon_2 -\alpha-1} , \label{eq:pdfJulius}
\end{equation}
and power law spectral density,
\begin{equation}
S(f) \sim \frac{1}{f^\beta}, \quad \beta = 1 + \frac{\varepsilon_2 + \alpha - 2}{1+\alpha} . \label{eq:specJulius}
\end{equation}
By using explicit form of
spectral density, Eq. (\ref{eq:specJulius}), we have reproduced
$1/f$ noise in three distinct cases of Eq. (\ref{eq:sdey}) (see
Fig. \ref{fig:compJulius}).

\begin{figure}
    \centering
    \includegraphics[width=0.4\textwidth]{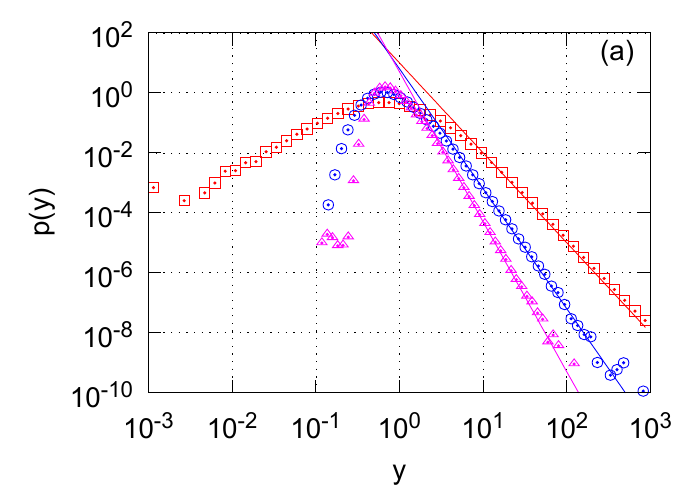}
    \hspace{0.1\textwidth}
    \includegraphics[width=0.4\textwidth]{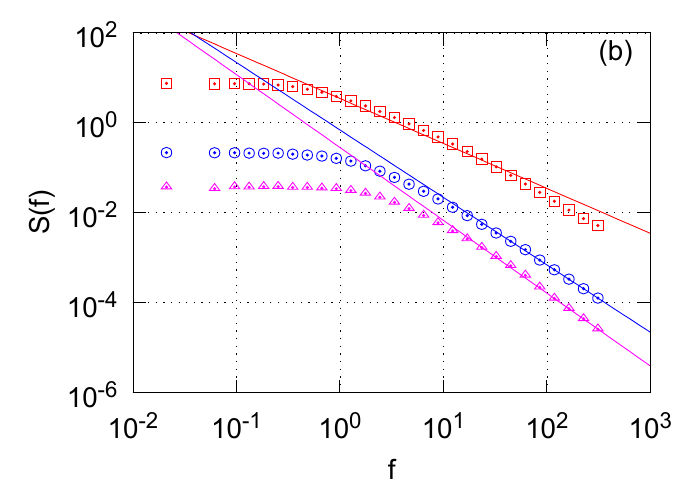}
    \caption{Numerically calculated PDF (a) and power spectral density (b) of $y$ defined by Eq. (\ref{eq:sdey}) with linear drift function in three distinct cases, $\alpha=0$ (red squares), $\alpha=1$ (blue circles) and $\alpha=2$ (magenta triangles). Other model parameters were set as follows: $\varepsilon_1=\varepsilon_2 = 2$. Solid curves are analytical fits, Eqs. (\ref{eq:pdfcev}) and (\ref{eq:speccev}), for modelic results: (a) $\lambda=3$ (red squares), $\lambda=4$ (blue circles), $\lambda=5$ (magenta triangles), (b) $\beta=1$ (red squares), $\beta=1.5$ (blue circles), $\beta=1.66$ (magenta triangles).}
    \label{fig:compCEV}
\end{figure}

If we linearize drift function of Eq. (\ref{eq:sdey}) with the respect to return (i.e. set $\varepsilon_2 = 2$), we obtain stochastic differential equation (once again in the limit $y \gg 1$),
\begin{equation}
\rmd y = \varepsilon_1 y \rmd t_s + \sqrt{2 y^{3+\alpha}} \rmd W_s . \label{eq:sdeylarge2}
\end{equation}
which is similar to the generalized CEV process
considered in \cite{reimann2011}:
\begin{equation}
\rmd y = a y \rmd t + b y^\eta \rmd W , \label{eq:cev}
\end{equation}
which is noted to be a special case of Eq.
(\ref{eq:JuliusBendra}), when exponential restriction of diffusion
is applied. Though comparison with this special case is important on its
own as this equation generalizes some stochastic models used in
risk management. Theoretical prediction of PDF
and spectral density for $y$ defined by Eq. (\ref{eq:cev}), is given
by \cite{reimann2011}:
\begin{eqnarray}
& p(y) \sim y^{-3-\alpha} , \label{eq:pdfcev}\\
& S(f) \sim \frac{1}{f^\beta}, \quad \beta=1+\frac{\alpha}{1+\alpha} , \label{eq:speccev}
\end{eqnarray}
where we have used previously set relation between model parameters $\eta$ and $\alpha$, Eq. (\ref{eq:etadef}). As one can see in Fig. \ref{fig:compCEV} these predictions are
also correct for $y$ defined by Eq. (\ref{eq:sdey}), in case of
$\varepsilon_2 = 2$.

Financial market time series are also known to exhibit interesting scaling behavior, namely they are known to be multifractal \cite{bouchaud2004, peters1994}. Thus a succesful model of financial market should generate multifractal time series. It is known that formaly constructed, via inverse Fourier transform, time series are not multifractal despite having proper spectral density, namely $S(f) \sim 1/f^\beta$, while the point process based models were shown to exhibit both power law spectral density and multifractality \cite{kaulakys2006}. As Eq. (\ref{eq:JuliusBendra}) was primarily derived using the very same point process model we expect its and therefore Eq. (\ref{eq:sdey}) solutions also to exhibit multifractality.

\begin{figure}
    \centering
    \includegraphics[width=0.4\textwidth]{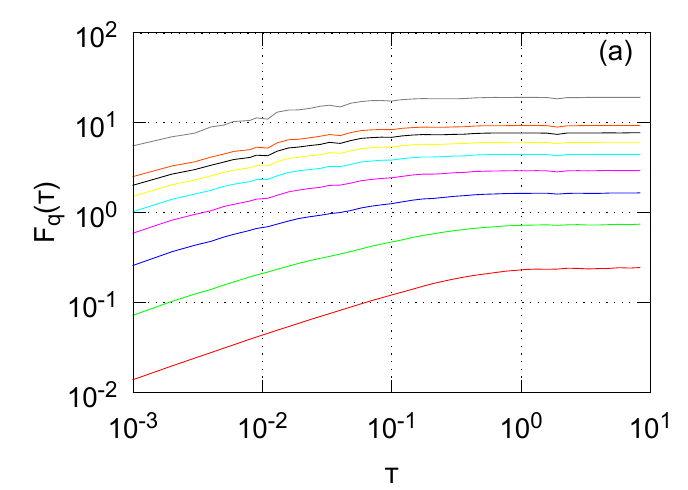}
    \hspace{0.1\textwidth}
    \includegraphics[width=0.4\textwidth]{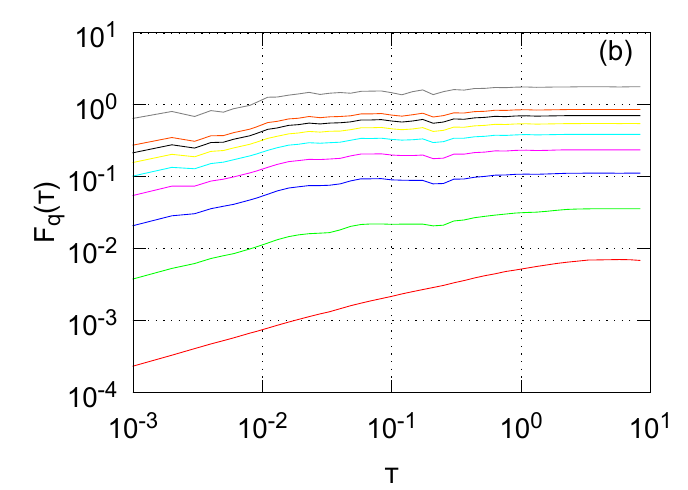}\\
    \includegraphics[width=0.4\textwidth]{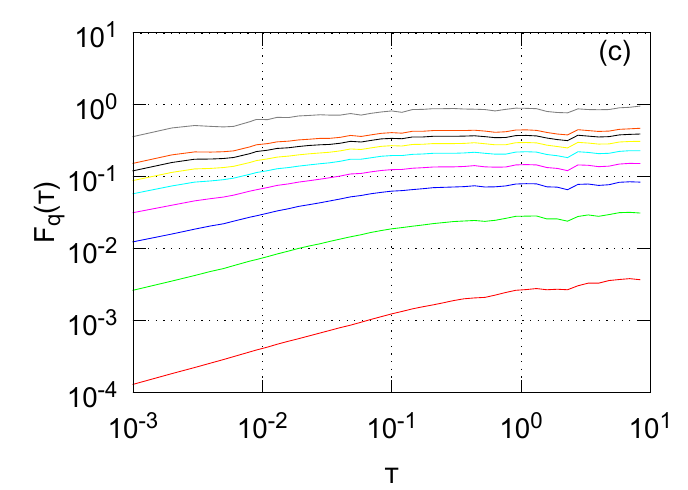}
    \hspace{0.1\textwidth}
    \includegraphics[width=0.4\textwidth]{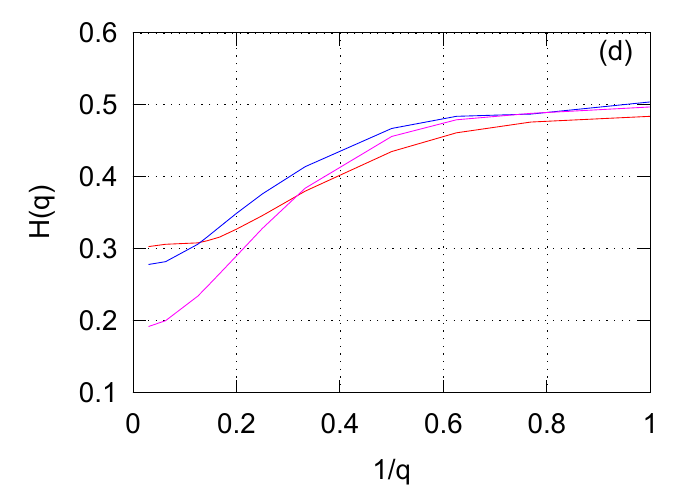}
    \caption{Generalized height-height correlation functions, $F_q(\tau)$, versus lag, $\tau$, for the $y$ time series obtained by solving  Eq. (\ref{eq:sdey}) in three distinct cases, $\alpha=0$ (a), $\alpha=1$ (b), $\alpha=2$ (c). Subfigure (d) is a plot of Hurst exponent, $H(q)$, versus its order, $q$, - red curve corresponds to $\alpha=0$ (or subfigure (a)), blue curve to $\alpha=1$ (or (b)) and magenta curve to $\alpha=2$ (or (c)). Other model parameters were set as follows: $\varepsilon_1=0$, $\varepsilon_2 = 2-\alpha$. In all three cases we have used single realization of million points to obtain $F_q(\tau)$. Nine orders of correlation, $q=1, 2, \dots, 8, 16$, are ploted in each of subfigures.}
    \label{fig:multifrac}
\end{figure}

As one can see in Fig. \ref{fig:multifrac} time series obtained by solving Eq. (\ref{eq:sdey}) are indeed multifractal. Power law slopes, characterized by Hurst exponent, $H(q)$, of generalized height-height correlation function, which is defined as
\begin{equation}
F_q(\tau) = \langle | y(t+\tau) - y(t) |^q \rangle^{1/q} ,
\end{equation}
where angular brackets, $\langle \dots \rangle$, denote time average, differ for different orders of correlation, $q$, in all three different cases of $\alpha$. Note that for all orders of correlation Hurst exponent values are smaller than $0.5$. This result is expected as mutlifractal time series exhibiting pink noise, $S(f) \sim 1/f^\beta \,,\,0<\beta<2$, are known to have Hurst exponent smaller than $0.5$ \cite{peters1994}.

\section{Conclusions and future work}
\label{sec:conclusion}
In this contribution we have started from a simple agent based model and obtained stochastic model for absolute return, Eq. (\ref{eq:sdey}). We found that obtained macroscopic model of absolute return is similar, in the limit of large values, $y \gg 1$, to the empirically derived stochastic models considered in \cite{reimann2011, ruseckas2010}. We have backed discussed analytical insights numerically - analytical predictions obtained in \cite{reimann2011, ruseckas2010} fit modelic results very well.

As considered stochastic model, Eq. (\ref{eq:sdey}), was obtained from Kirman's agent based model, Kirman dynamics can be seen as a microscopic explanation of some very general non-linear stochastic models and, thus, statistical properties, namely long-range memory and fat tails, observed in actual financial markets. The exponents of power law statistics, $\lambda$ and $\beta$, can be adjusted by introducing feedback between the macroscopic system state, $y$, and the rate of microscopic events $1/\tau(y)$, where $\tau(y) = y^{-\alpha}$.

Basing ourselves on this simple model we can further work on replication of more sophisticated statistical properties of return, namely fractured spectral density, and modeling of trading activity. We also think that it is possible to apply Kirman model more broadly - ideas underlying such birth-death processes are very well spread over different fields (ex. marketing \cite{daniunas2011}, social dynamic modeling \cite{ausloos2011}).

\section{Acknowledgment}
The work was done in the frame of European program COST Action
MP0801.

\bibliographystyle{elsarticle-num}
\bibliography{microreasoning}

\end{document}